# STRUCTURE OF SUPEROXIDE REDUCTASE BOUND TO FERROCYANIDE AND ACTIVE SITE EXPANSION UPON X-RAY INDUCED PHOTO-REDUCTION.

**Authors:**

Virgile Adam[1,2], Antoine Royant[1,3], Vincent Nivière[4], Fernando P. Molina-Heredia[4] and Dominique Bourgeois[1,2].

**Affiliations:**

[1] *LCCP, UMR 5075, IBS-CEA/CNRS/Université J. Fourier, 41 avenue Jules Horowitz, 38027 Grenoble, Cedex 1, France*

[2] *ESRF, BP 220, 38043 Grenoble Cedex, France*

[3] *EMBL, 6, rue Jules Horowitz, BP181, 38042 Grenoble Cedex 9, France*

[4] *CBCRB, UMR 5047, DRDC-CEA/CNRS/Université J. Fourier, 17 avenue des Martyrs, 38054 Grenoble, Cedex 9, France*

**Running Title:**

Structure of SOR bound to ferrocyanide.

**Corresponding author:**

**D. Bourgeois** (e-mail: bourgeoi@lccp.ibs.fr)

LCCP, UMR 9015, IBS, 41 rue Jules Horowitz, 38027 Grenoble Cedex 1, France.

Tel: +33 (0)4 38 78 96 44      Fax: +33 (0)4 38 78 51 22

**Abbreviations:**

SOR: superoxide reductase

SOD: superoxide dismutase

FTIR: Fourier transform infrared spectroscopy

ESI: electron spray ionization

EPR: electron paramagnetic resonance

Rmsd: root-mean-square deviation

**Data deposition:**

Coordinates and structure factor amplitudes of the structures: $SOR_{E47A, Fe(II)}$, $SOR_{E47A, Fe(III)}$-$Fe(II)(CN)_6$, and $SOR_{E47A, Fe(II)}$-$Fe(II)(CN)_6$ have been deposited in the RCSB Protein Data Bank with identification codes XXX, YYY and ZZZ, respectively.

## SUMMARY

Some sulphate-reducing and microaerophilic bacteria rely on the enzyme superoxide reductase (SOR) to eliminate the toxic superoxide anion radical ($O_2^{\bullet-}$). SOR catalyses the one-electron reduction of $O_2^{\bullet-}$ to hydrogen peroxide at a non-heme ferrous iron center. The structures of *Desulfoarculus baarsii* SOR (mutant E47A) alone and in complex with ferrocyanide were solved to 1.15 and 1.7 Å resolution, respectively. The latter structure, the first ever reported of a complex between ferrocyanide and a protein, reveals that this organo-metallic compound entirely plugs the SOR active site, coordinating the active iron through a bent cyano bridge. The subtle structural differences between the mixed-valence and the fully-reduced SOR-ferrocyanide adducts were investigated by taking advantage of the photo-electrons induced by X-rays. The results reveal that photo-reduction from Fe(III) to Fe(II) of the iron center, a very rapid process under a powerful synchrotron beam, induces an expansion of the SOR active site.

## INTRODUCTION

The superoxide anion radical $O_2^{\bullet-}$, the univalent reduction product of molecular oxygen, is highly toxic for living organisms (Imlay, 2003). For years, the only enzymatic system known to catalyze elimination of superoxide was superoxide dismutase (SOD) (Fridovich, 1995). Recently, a new mechanism of cellular defense against oxidative stress was discovered, involving the non-heme iron protein superoxide reductase (SOR). SOR catalyzes the one-electron reduction of $O_2^{\bullet-}$ to hydrogen peroxide at a nearly diffusion-controlled rate (Jenney et al., 1999, Lombard et al., 2000a, Niviere et al., 2004b):

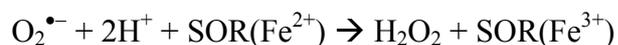

$$O_2^{\bullet-} + 2H^+ + SOR(Fe^{2+}) \rightarrow H_2O_2 + SOR(Fe^{3+})$$

In contrast to SOD, SOR does not catalyze oxygen production, and up to now it has been only found in some anaerobic or microaerophilic bacteria or archae (Niviere et al., 2004b). In these organisms, it has been shown to play a fundamental role in case of oxidative stress provoked by exposure to adventitious molecular oxygen (Emerson et al., 2003b, Fournier et al., 2003, Pianzzola et al., 1996). Understanding the catalytic mechanism engineered by SOR is of fundamental interest, and is also needed because the enzyme is potentially a target of considerable pharmacological relevance, for example in the case of the pathogenic bacterium *Treponema pallidum*, responsible for syphilis (Fraser et al., 1998).

The three-dimensional structures of *Desulfovibrio desulfuricans* and *Pyrococcus furiosus* SORs have revealed a catalytic domain formed by a seven stranded β-sandwich displaying an immunoglobulin-like fold (Coelho et al., 1997, Yeh et al., 2000). Catalysis occurs at a ferrous iron center (named Center II) (Lombard et al., 2000a) adopting a square pyramidal coordination to four histidines in equatorial positions and a cysteine in axial position. The sixth coordination site, where $O_2^{\bullet-}$ is thought to bind, remains vacant in the reduced state and is occupied by a strictly conserved glutamate in the oxidized state (Berthomieu et al., 2002, Clay et al., 2002, Yeh et al., 2000). Three classes of SORs have been described so far, all of which display a highly similar active site. *D. desulfuricans*, *Desulfovibrio vulgaris* and *Desulfoarculus baarsii* SORs are representatives of class I, in which the catalytic domain is linked to a small N-terminal domain structurally similar to desulforedoxin (Archer et al., 1995) and containing an additional iron center, named Center I (Coelho et al., 1997). The mononuclear iron(III) of Center I is ligated by four cysteines in a distorted tetrahedral arrangement, of rubredoxin-type. SORs of class II, as *P. furiosus* SOR, are characterized by the absence of the additional N-terminal domain (Yeh et al., 2000). In class III, this domain is present, but center I is absent, such as in *T. pallidum* SOR (Jovanovic et al., 2000, Lombard et al., 2000b). In fact, when present, a functional role of Center I has not been established (Emerson et al., 2003a, Lombard et al., 2000a).

To date, the formation of SOR complexes with ligands or substrate analogs has been investigated by various spectroscopic techniques, but a structural characterization of these species at the atomic level has been missing. Formation of SOR adducts with azide, cyanide and nitric oxide have been studied (Clay et al., 2002, Clay et al., 2003a) and were found consistent with an inner-sphere catalytic mechanism involving the formation of iron(III)-(hydro)peroxo intermediate species (Abreu et al., 2001, Emerson et al., 2002, Lombard et al., 2001, Mathe et al., 2002, Niviere et al., 2004a, Silaghi-Dumitrescu et al., 2003). The formation of a stable cyano-bridged dinuclear iron cluster following oxidation of SOR with hexacyanoferrate(III) (also referred to as ferricyanide or $Fe(III)(CN)_6$) was first detected by EPR studies (Clay et al., 2002) and was subsequently investigated by FTIR, ESI mass spectrometry and near-IR variable-temperature magnetic circular dichroism (Auchere et al., 2003, Clay et al., 2003b). Analysis of FTIR stretching bands suggested the existence of a $Fe(III)-NC-Fe(II)(CN)_5$ complex with the bridging cyanide adopting a bent geometry (Auchere et al., 2003).

Here, we have solved the three-dimensional (3D) structures of *D. baarsii* SOR E47A mutant, alone and in complex with hexacyanoferrate(II) (also referred to as ferrocyanide or $Fe(II)(CN)_6$), to 1.15 and 1.7 Å resolution, respectively. The E47A mutant was initially chosen because of its ability to stabilize a (hydro)peroxo intermediate (Mathe et al., 2002), opening future possibilities to form such an entity in the crystal. Although the release of $H_2O_2$ may be assisted by the ligation of Glu 47 to the oxidized active iron (Mathe et al., 2002, Niviere et al., 2004a, Yeh et al., 2000), it has been shown that the mutation of this residue to alanine has no effect on the rate of the reaction of SOR with superoxide (Abreu et al., 2001, Emerson et al., 2002, Lombard et al., 2001, Niviere et al., 2004a), suggesting no significant alteration of the 3D structure of SOR.

The structures of the mixed-valence and fully reduced complexes ($SOR_{Fe(III)}$-$Fe(II)(CN)_6$ and $SOR_{Fe(II)}$-$Fe(II)(CN)_6$, respectively) are the first ever reported of a protein bound to ferrocyanide. Both structures were obtained from a single crystal by taking advantage of the photo-electrons induced by the X-ray beam (Berglund et al., 2002). Photo-reduction of the crystalline enzyme was monitored online by absorption microspectrophotometry and the associated subtle structural changes could be observed.

Our data bear interest for the understanding of the SOR catalytic mechanism and the rational design of drugs targeting the enzyme. They show that substrate binding to SOR involves a significant electrostatic component, reveal a new type of cyano-bridged dinuclear iron cluster, and suggest a subtle volume expansion of the SOR active site upon reduction.

## RESULTS

### Structure of *D. baarsii* SOR (E47A mutant, 1.15 Å resolution)

*D. baarsii* SOR (Table 1, Figure 1) E47A is a homodimer whose overall fold closely resembles the structure of *D. desulfuricans* SOR (rmsd on backbone atoms is 0.55 Å, to be compared with 0.20 Å between the two monomers of the *D. baarsii* enzyme). Both monomers (referred to as monomers A and B, respectively) are composed of the two iron-containing domains typical of Class I SORs. Data collection statistics and refinement parameters are shown in Table 1.

In the reduced state, the E47A mutation is unlikely to induce any significant reorganization of the active center as compared to the wild-type enzyme. This is indicated by the observed superposition of this center in our structure, onto the wild-type *D. desulfuricans* structure (rmsd's on backbone and side-chain atoms less than 8 Å away from the active iron are 0.31 Å and 1.21 Å, respectively; 17 residues are concerned, displaying 82 % sequence identity and 100 % similarity between the two structures). Hence, the high resolution of our data provides a highly precise representation of the active site geometry (Table 2). The active iron is found 0.5 Å out of the equatorial plane formed by the four histidine ligands, in the direction of Cys116. Interestingly a strong electron density feature is observed in the putative substrate binding cavity (6.0 $\sigma$ in the initial $F_o$-$F_c$ difference electron density computed before modeling), at a distance of 4.2 Å from the iron (Figure 2). A similar density was noticed in the case of *D. desulfuricans* SOR, 4.9 Å from the iron, but was not assigned, probably because of the lower resolution (Coelho et al., 1997). We attribute this density to a chloride ion on the basis of (*i*) the absence of residual $F_o$-$F_c$ difference electron density when this ion is included in the model, (*ii*) the value of the B-factor refined for chloride (19.6 Å$^2$), which is similar to those of neighboring atoms (15.7 Å$^2$ on average for residues less than 5 Å away from the anion) and (*iii*) the fact that 0.1 M calcium chloride was present in the crystallization medium. The chloride ion remains partially solvated by four water molecules situated at a mean distance of ~ 3.1 Å, two of which are hydrogen bonded to the protein. In particular, Wat113 is strongly bound to Lys48, at a distance of 2.8 Å from atom $N_\xi$, correlating with the essential role of this strictly conserved residue which sticks out of the protein surface and serves as the primary attractor for the anionic superoxide substrate (Lombard et al., 2001).

**The SOR-Fe(CN)$_6$ complex: single crystal microspectrophotometry and monitoring of X-ray induced photoreduction.**

Oxidation of SOR by potassium ferricyanide has been suggested to produce a stable ferrocyanide adduct with the enzyme active site (Auchere et al., 2003, Clay et al., 2002). As this adduct is expected to display a 6-coordinated active center iron, possibly analogous to the octahedral geometry adopted upon substrate binding, we set out to co-crystallize the *D. baarsii* enzyme with ferricyanide. Single-crystal microspectrophotometry (Bourgeois et al., 2002) revealed that Fe(III)(CN)$_6$ oxidizes the active center iron as assessed by the appearance of a new band at ~ 650 nm in the absorption spectrum (Figure 3a). The absence of a band at ~ 420 nm after washing the crystal suggests that ferricyanide has been reduced to ferrocyanide. In addition, the absorbance bands from Center I (in the 400-570 nm region) are weaker than in the absence of Fe(III)(CN)$_6$, suggesting a partial depletion in the iron content of this center caused by the treatment with ferricyanide. Such a depletion is not expected to be of functional importance, as an engineered *D. vulgaris* SOR lacking Center I has been shown to retain full catalytic activity (Emerson et al., 2003a).

Exposure for 90 s under the X-ray beam of ESRF station ID14-EH4 resulted in the complete disappearance of all absorption bands assigned to the oxidized states of both iron Centers I and II (Figure 3b). We assign this observation to X-ray induced photo-reduction (Chance et al., 1980, Karlsson et al., 2000). As a result, the exact valence state of the SOR-Fe(CN)$_6$ complex in a X-ray structure could not be established unambiguously in the absence of a tight control of the X-ray dose delivered to the crystal.

Online microspectrophotometry (Ravelli et al., unpublished results, see also Sakai et al., 2002) was used to quantitatively evaluate the reduction of SOR crystals by X-ray induced photo-electrons. Time-resolved difference absorption spectra showed a rapid decay of the absorption band at 650 nm (Figure 4a, a movie is available as Supplementary Material) upon exposure to a (0.1 x 0.1) mm$^2$ X-ray beam at 0.92 Å delivering ~ 5x10$^{11}$ photons s$^{-1}$. Assuming that this decay accurately describes the photo-reduction kinetics of Center II, more than 25% of the active center irons are reduced by the X-ray beam in about 3 seconds, long before any radiation damage becomes apparent on diffraction patterns.

A special data collection strategy was therefore required to solve the structure of the SOR-Fe(CN)$_6$ complex in well-defined redox states. Using the recent method of Berglund *et al.* (Berglund et al., 2002) we collected two "composite" data sets from a single rod-shaped crystal. In the first composite data set (referred to as the "LD" data set), a low X-ray dose was deposited into the sample, whereas in the second one (referred to as the "HD" data set), a higher dose was deposited. From the LD and HD data sets, models of the *D. baarsii* SOR in complex with ferrocyanide could be refined to 1.7 Å resolution, that correspond to the oxidized and reduced states of the active center iron, respectively. Furthermore, the subtle structural changes induced by photo-reduction of this center could be observed with minimum systematic errors.

**Structures of the mixed-valence and reduced SOR-ferrocyanide complexes (E47A mutant, 1.7 Å resolution).**

The structure of the mixed-valence $SOR_{Fe(III)}$-Fe(II)(CN)$_6$ complex reveals that ferrocyanide entirely plugs the enzyme active site and shows an octahedral geometry of the active center iron (Figure 5). One apical cyanide bridges the two irons (distance $N_{1, Fer}$-Fe(III)$_{SOR}$ ≈ 2.36 Å), whereas the other sticks out of the protein. Coordination of Fe(CN)$_6$ to the SOR iron occurs with a bent geometry (Table 2), due to asymmetric anchoring of the equatorial cyanides on the SOR active site (Figure 5d). Indeed, ferrocyanide is interacting with Ala45 and Lys48, with hydrogen bonds between $N_5$ and $N_2$ of Fe(CN)$_6$ to O of Ala45 and $N_\zeta$ of Lys48, respectively, and van der Waals interactions between $N_3$ of Fe(CN)$_6$ and $C_\beta$ of Ala45 and $C_\delta$ of Lys48. As a result of the bent geometry (angle Fe(II)-$N_1$-Fe(III)$_{SOR}$ ≈ 142°), ferrocyanide establishes van der Waals contacts with His119 and His75 through the fourth cyanide group nitrogen. Binding of Fe(CN)$_6$ also results in pulling the active center iron back into the equatorial plane (distance Fe – equatorial plane ≈ 0.15 Å), resulting in the $S_{\gamma, Cys116}$–Fe coordination bond being stretched by ~ 0.3 Å (Table 2). The distances between the active iron and the equatorial histidines are not significantly modified.

The presence of ferricyanide in the crystallization medium did not induce noticeable structural changes of SOR other than at the active site. However, a depletion of the iron content in Center I is observed (refined occupancy for the iron ≈ 40%), confirming the predictions made from microspectrophotometry. As a consequence, Cys13 and Cys30 as well as Cys10 and Cys29 form disulfide bridges.

The structure of the SOR-Fe(CN)$_6$ complex in the fully reduced state shows small but significant differences from that of the mixed-valence complex. X-ray induced photo-reduction of Center II at 100 K induced structural modifications of the active iron coordination pattern, as revealed by computation of electron density difference maps (Figure 6a). These modifications are almost identical in both monomers. First, an outward motion of ferrocyanide is observed, suggesting that the dinuclear iron cluster formed by the active center iron and $Fe(CN)_6$ elongates upon reduction. Second, a positive feature appears at the level of the active center iron which is not compensated by any negative feature. This observation suggests the absence of a net motion of the iron upon reduction, and is consistent with either a reduced thermal agitation, or, more likely, with an increased net partial negative charge. Third, a positive density, again not compensated by any negative feature, is observed behind $S_\gamma$ of Cys116, suggesting a stretching of the $S_\gamma$-Fe coordination bond and either a reduced thermal agitation or an increased net partial negative charge of the cysteinyl sulfur atom. It should be noted that the observed features in Figure 6a are very unlikely to result from Fourier truncation errors around a heavy atom, as no overwhelming density is observed in the difference electron density map.

At the level of Center I, the disulfide bridges formed upon iron depletion in the SOR-$Fe(CN)_6$ complex clearly break up upon photo-reduction (Figure 6b). This observation is expected from previous studies on the radiation sensitivity of disulfide bridges (Alphey et al., 2003, Weik et al., 2000), and corroborates the development of absorption below 450 nm observed by online microspectrophotometry (Weik et al., 2002). In the present case, it provides us with an internal control of the quality of the composite data sets.

Model refinement of the mixed-valence and fully reduced structures confirmed these qualitative observations. Upon reduction, the $Fe_{SOR}$-$S_{\gamma, C116}$ and the $Fe_{SOR}$-$N_{1, Fer}$ bonding distances elongate by ~ 0.06 Å and ~ 0.07 Å, respectively. Additionally, some of the equatorial histidines move away from the active iron. His119, the most solvent-exposed equatorial ligand, is displaced by ~ 0.05 Å, whereas His49, the most buried one, shows no movement. However, the significance of these latter motions is not supported by the presence of peaks above noise level in the Fourier difference map.

Overall, the octahedral volume formed by the atoms coordinating the active center iron expands by ~ 4.9 % upon reduction (Table 2).

**Catalytic activity of *D. baarsii* SOR in the presence of ferrocyanide.**

The structures of the SOR-ferrocyanide complex suggest that ferrocyanide, by plugging the active site, could act as a potent inhibitor of the SOR activity. The effect of ferrocyanide on SOR activity was investigated in solution. The kinetics of the oxidation of Center II by $O2^{\bullet-}$, generated by the xanthine-xanthine oxidase system was followed by spectrophotometry at 640 nm. As shown in Figure 7, below about 2 molar equivalents of ferrocyanide with respect to SOR, the velocity of SOR oxidation by superoxide decreases dramatically with increasing amount of ferrocyanide present in the assay. However, above 2-5 equivalents of ferrocyanide, the velocity of SOR oxidation becomes almost ferrocyanide-independent, with a value of about 20% of that in the absence of ferrocyanide. Identical results are obtained in the cases of the wild-type SOR and the E47A mutant (Figure 7).

**DISCUSSION**

The X-ray crystallographic structures of superoxide reductases from *D. desulfuricans* (Coelho et al., 1997) and *P. furiosus* (Yeh et al., 2000) have revealed a highly unusual non-heme iron active site, adopting a square pyramidal geometry and being particularly well exposed to the solvent. In this work, we have demonstrated that the unique properties of the SOR active site confer to the enzyme the ability to form a tight complex with ferrocyanide. The structure of this complex is the first ever reported of a protein bound to this organo-metallic compound, and reveals a new type of cyano-bridged dinuclear iron cluster.

Spectroscopic studies have previously suggested that oxidation of SOR by ferricyanide produces a stable ferrocyanide adduct with the iron active site (Auchere et al., 2003, Clay et al., 2002). FTIR analysis following oxidation of *D. vulgaris* and *T. pallidum* SOR by ferricyanide have showed a characteristic C-N stretching band at 2095 cm$^{-1}$ suggestive of a cyano-bridged dinuclear iron cluster adopting a bent geometry (Auchere et al., 2003). Such a geometry is clearly observed here: ferrocyanide interacts with two of the liganding histidines, His75 and His 119, through van der Waals contacts, whereas the side chains from Ala45 and Lys48 prevent it from contacting the two other histidines (Figure 5b).

The remarkable plugging of ferrocyanide results from electrostatic complementarity between $Fe(CN)_6$ and the SOR active site (Figure 5a, b), but also from steric complementarity. Crevices at the enzyme surface accommodate the equatorial cyanide moieties of $Fe(CN)_6$ (Figure 5a, b). Moreover, the fourfold symmetry of these equatorial cyanides matches the spatial distribution of the four equatorial histidines of Center II, resulting in an interleaved stacking that enhances the stabilization of the complex (Figure 5c).

The above-mentioned interactions between SOR and $Fe(CN)_6$ are compatible with the establishment of a coordination bond between the active center iron and the nitrogen atom of the bridging cyanide. However, the protein environment constrains the distance between these two atoms (2.36 Å in the mixed-valence state) to be greater than the one measured on the chemically and structurally related organo-metallic compound Prussian Blue (2.03 Å, (Buser et al., 1977)).

Inspection of our structures of reduced SOR with a chloride ion and of the SOR-$Fe(CN)_6$ complexes indicate that substrate binding to SOR involves a significant electrostatic component involving Lys48. It has been observed that a SOR mutant where this residue is replaced by isoleucine leads to a ~30 fold decrease in the kinetics of binding of superoxide to the enzyme (Lombard et al., 2001). Our structures show that Lys48 sticks out of the protein surface and interacts with either $Fe(CN)_6$ or chloride in the absence of $Fe(CN)_6$, supporting the view that its physiological role consists in grabbing anionic substrate molecules to funnel them towards the active site (Lombard et al., 2001).

Binding of chloride to an iron center is an unusual finding, which has only been reported in the case of *T. zostericola* myohemerythrin, in which the anion directly coordinates a diiron center (Martins et al., 1997). Direct coordination does not occur in SOR, as chloride remains trapped to the protein's surface through four water molecules of its hydration shell. A practical image would be the one of a parachutist suspended to a tree by his parachute, unable to touch the ground. Given the geometry of the active site, direct coordination to the iron would require almost complete desolvation of chloride, which would be of too high free-energy cost. Hence, our structure illustrates the energy balance between electrostatic interaction and ion desolvation. Interestingly, the binding mode of chloride strikingly matches the one of ferrocyanide (Figure 5d and e). Each of the four water molecules "corresponds" to an equatorial cyanide group, and establishes the very same interactions with neighboring atoms from the protein. This supports the idea that anions exhibiting affinity for SOR interact in a common manner with the active site. Besides, as the SOR-ferrocyanide adduct displays a 6-coordinated active center iron, it would be tempting to propose that the structure of the complex mimics the octahedral geometry adopted upon substrate binding, with superoxide occupying a similar position as the bridging cyanide. However, such a conclusion may only be drawn with the greatest care.

Cyanide has been mentioned to be a weak inhibitor of SOR activity (reduction of the enzymatic activity by a factor of 7 when 50 equivalents of CN⁻ are added), (personal communication from M. K. Johnson and M. W. W. Adams reported in Shearer et al., PNAS, 2003). With ferrocyanide, only two equivalents are sufficient to reduce the enzymatic activity by a factor of 5 (Figure 7). This suggests that the SOR-Fe(CN)$_6$ complex is tightly formed in solution. In addition, since the wild-type SOR behaves exactly as the E47A mutant, it appears that Glu47 does not interfere with binding of ferrocyanide. These findings are consistent with the striking electrostatic and steric complementarity between Fe(CN)$_6$ and the SOR active site (Figure 5), which are likely to largely increase the binding affinity of Fe(CN)$_6$ relative to CN⁻. On the other hand, because ferrocyanide entirely plugs the SOR active site and prevents superoxide to approach the active iron, it was expected to act as a more efficient inhibitor of the enzyme. The remaining SOR activity with superoxide which is observed in the presence of an excess of Fe(CN)$_6$ appears then rather surprising. This activity might be associated with an alternative $O_2^{\bullet-}$ reduction mechanism involving the still redox active ferrocyanide. Nevertheless, the structure and the stability of the SOR-Fe(CN)$_6$ complex constitute a basis of how drugs specific to SOR could be designed. Others hexacyanate complexes, with metals expected to be redox inert in biological media, might exibit a similar binding capacity to the SOR active site, without any residual activity. Alternative hexacyanate complexes formed with Ru, Co and Pt could for example be considered.

During data collection, the redox state of the crystalline SOR-ferrocyanide complex could be monitored online by time-resolved absorption microspectrophotometry. Photo-reduction of the active center iron by a highly brilliant synchrotron X-ray beam was shown to occur within a few seconds. Photo-reduction of metal centers in crystalline enzymes has been reported in several cases (Berglund et al., 2002, Chance et al., 1980, Karlsson et al., 2000, Logan et al., 1996, Schlichting et al., 2000, Sjogren et al., 2001) but a quantitative evaluation of the rate of the process was not carried out to date. The rate observed in the case of SOR is surprisingly rapid, but is consistent with X-ray absorption spectroscopic studies on the *P. furiosus* enzyme having shown that the active site of SOR is highly susceptible to photoreduction by X-rays (Clay et al., 2002). This susceptibility probably relates to the accessibility of the solvent to the active center iron.

Elegant ideas have recently been proposed to turn radiation damage into an ally (Berglund et al., 2002, Ravelli et al., 2003). Here, we took advantage of X-ray induced photo-electrons to study the structural changes associated with the reduction of the active center iron in the SOR-$Fe(CN)_6$ complex. Comparison of the mixed-valence and fully reduced structures shows a volume expansion ($\sim 4.9$ **%**) of the active site upon reduction at 100 K (Table 2). Such an expansion is expected, as the electrostatic interactions between the net positive charge carried by the active iron and the six lone electron pairs provided by the ligands decrease upon reduction. However, a direct experimental evidence, as presented here, was never provided.

The main contributors to the expansion are the axial ligands of the active iron, with a backward motion of ferrocyanide and a stretching of the $Fe-S_{\gamma, Cys116}$ bonding distance by more than 0.06 Å. Cys116 appears as the most flexible ligand of the active iron. Tuning of the SOR active site to accommodate $Fe(CN)_6$ involves a lengthening of the $Fe-S_{\gamma, Cys116}$ distance (increasing by 12 % relative to the apo structure, Table 2), suggesting a weakening of the $Fe-S_\gamma$ bond upon $Fe(CN)_6$ complexation, and in agreement with recent resonance Raman studies showing a downshift of the $Fe-S_\gamma$ stretching modes in the complex (C. Mathé, et al., submitted). Upon reduction, a further stretching by 2.3 % is observed. These observations are in line with the proposal that the $Fe-S_\gamma$ p$\pi$-d$\pi$ bonding interaction actively contributes to the active-site reactivity and may play a role in electron transfer (Clay et al., 2002). Furthermore, the positive feature, without any negative counterpart, observed at the level of Cys116 $S_\gamma$ in the electron difference density map of Figure 6a may reveal an increased partial negative charge on the cysteinyl sulfur atom. Such an interpretation would be in agreement with DFT studies on different Fe-S clusters having predicted that the net charge increase upon reduction is largely distributed to the sulphur atoms (Noodleman et al., 1992). DFT studies have been performed on SOR (Silaghi-Dumitrescu et al., 2003) and the technique could in principle be used to confirm this idea.

All displacements that lead to the overall expansion of the active site upon reduction are smaller than the estimated absolute coordinate errors of the structures (~ 0.15 Å). However, a "calculated" map of type ($F_{c, reduced} - F_{c, mixed-valence}$) $\exp(i\phi_{c, mixed-valence})$ essentially reproduces the features observed in Figure 6a and corroborates the consistency of our models (Figure S1, available as Supplementary Material). Considering the data collection strategy and the refinement protocol employed to obtain our final models, systematic errors are expected to be identical in the two structures. Therefore, errors in coordinate shifts between these structures are much smaller than the absolute coordinate error of each of them, giving significance to the subtle expansion observed. In addition, the motions of the axial ligands are real because they are clearly observed in experimental difference maps. This is not the case for the equatorial histidines, but in-plane translation of an imidazole ring results in a weak change of the electron density (Figure S1). Nevertheless, although the individual motions of the histidines are consistent with the steric constrains experienced by these residues, their significance remains questionable at this stage. In summary, photoreduction seems to induce an average backward motion of the equatorial ligands, in response to a decrease in their electrostatic interaction with the active iron.

Overall, a re-organization of the active center molecular orbitals upon a one-electron charge increase has been visualized. As the experiment was performed at cryo-temperature (100 K), below the dynamical transition (Parak et al., 1982), only small motions could be observed, and it may not be excluded that reduction at room temperature could induce more drastic conformational changes. However, the observations made here are of general interest for the study of redox-dependent structural changes of metal centers in proteins.

## CONCLUSION

The ability to selectively reduce the highly toxic superoxide anion confers unique structural properties to superoxide reductase. Our results show that SOR is designed to bind negatively charged molecules such as $O_2^{\bullet-}$. Moreover, the ability of the enzyme to make a complex with the organo-metallic compound hexacyanoferrate(II) is a remarkable feature. Indeed, despite the numerous protocols in which this molecule is used to oxidize metalloenzymes, the 3-dimensional structure of a cyano-bridged dinuclear metal center involving $Fe(CN)_6$ and a metalloenzyme has never been reported. In *D. baarsii* SOR, ferrocyanide is observed to coordinate axially the catalytic iron through a cyanide bridge adopting a bent geometry, and to completely block access to the SOR binding site. Therefore, it constitutes a model for rational drug design targeting the enzyme. The observation of a global expansion of the coordination pattern of the active center iron upon photo-reduction is of general interest for the study of redox-dependant structural changes in metalloenzymes. Lengthening of the iron sulphur bonding interaction and possible charge transfer to the sulphur atom upon reduction suggest that this interaction actively contributes to the reactivity of the SOR active-site.

## EXPERIMENTAL PROCEDURES

### Purification and crystallization of SOR

Wild type and E47A *D. baarsii* SORs were over-expressed in *E. coli* DH5α strains and purified as described (Lombard et al., 2001). For crystallization assays, an additional hydrophobic chromatography column (phenyl Superose, Pharmacia) was carried out after the anion exchange chromatography step.

E47A SOR crystals in the semi-reduced form were obtained using the hanging drop method by mixing 2μL of a crystallization solution (18% PEG 4000, 100 mM Tris buffer, pH 8.0 and 100 mM CaCl$_2$), with 2μL of a protein solution (1 mM SOR in 10 mM Tris buffer, pH 7.5). Needle-shaped, red-colored crystals of size ~ (0.5 x 0.1 x 0.1) mm$^3$ grew in about two weeks at 20°C, in the tetragonal space group I4$_1$22 with cell dimensions a = b = 123.61 Å, c = 73.02 Å, α = β = γ = 90º. The asymmetric unit comprises the SOR homodimer and the crystals contain 38% solvent forming large channels (~ 30 Å diameter). However, accessibility to the active site in both monomers is restricted, which probably explains why soaking experiments with hexacyanoferrate(III) failed.

Crystals of the complex between SOR and hexacyanoferrate(II) were obtained with the same crystallization solution to which 4 mM K$_3$Fe(III)(CN)$_6$ was added. Co-crystallization induced a change to an orthorhombic space group. Needle shaped, brown-colored crystals of size ~ (0.6-1 x 0.15 x 0.15) mm$^3$ grew in about one month at 20°C in the P2$_1$2$_1$2$_1$ space group with cell dimensions a = 42.13 Å, b = 67.65 Å, c = 82.19 Å, α = β = γ = 90º. The crystals contain 36% solvent.

For X-ray data collection and microspectrophotometry experiments, crystals were transferred to a cryoprotectant solution (5 μL of the crystallization solution to which 20% glycerol is added) for ~1 min prior to being flash cooled to 100 K in gaseous nitrogen with a nylon loop.

**Offline absorption microspectrophotometry.**

Absorption spectra on SOR crystals were collected using a Deuterium lamp (Oriel 63163) and an offline microspectrophotometer (Bourgeois et al., 2002). To check the effect of X-ray induced photoreduction, crystals were transferred with cryo-tongs and absorption spectra were collected before and after X-ray data collection on the same crystal (Figure 3).

**Online absorption microspectrophotometry.**

Absorption spectra on SOR crystals were collected online using a device recently installed on station ID14-EH4 at the ESRF (Ravelli *et al.*, in preparation). A 1 mm long, rod-shaped crystal was mounted at 100K and submitted to a $(0.1 \times 0.1)$ mm$^2$ X-ray beam at 0.92 Å delivering $\sim 5 \times 10^{11}$ photons s$^{-1}$. The part of the crystal volume exposed to X-rays was probed by a 25 μm diameter beam from a Xenon light source. Difference absorption spectra (from an initial spectrum recorded before exposure to X-rays) were collected upon X-ray irradiation at a frequency of $\sim 10$ Hz. The decay of the absorption band at $\sim 650$ nm (Figure 4b) displays a bi-exponential behaviour, with half-times of 6.5 sand 72 sec. Other features of interest were observed in addition to the rapid decay of the $\sim 650$ nm band. The fastest process was the appearance of a broad absorption band centered at $\sim 540$ nm, tentatively assigned to the presence of trapped hydrated electrons within the sample (Ershov et al., 1968). A major absorption increase was also observed below 450 nm that corroborates bleaching of the crystal observed visually (Figure 4, inset), and that can be attributed to reduction of disulfide bridges within the sample.

**Data collection and structure solution**

Data collections were performed on beamlines BM30A, ID29, ID14-EH2 and ID14-EH4 of the ESRF, Grenoble, France. The structure of reduced E47A SOR was solved by molecular replacement using the structure of *D. desulfuricans* (accession code: 1DFX) to which all hetero atoms were removed, as a starting model. Model refinement was initially performed with CNS (Brunger et al., 1998) to 1.5 Å resolution, and then extended to 1.15 Å with Refmac (Murshudov et al., 1997). Individual atomic thermal parameters were refined anisotropically. Dual conformations were modeled for some residues or atoms, where necessary. No restraints were applied on the metal ions. A final round of conjugate gradient minimization was carried out with Shelx (Sheldrick, 1997 ) and a ultimate full-matrix least square refinement without restraints yielded estimated standard deviations on all individual atomic positions (Table 2).

To solve the structure of the SOR-Fe(CN)$_6$ complex, an initial model was derived by positioning the E47A SOR structure (with hetero atoms removed) into the new unit cell by rigid body refinement, and by placing manually ferrocyanide molecules into the electron density using O (Jones et al., 1991). Ferrocyanide was built as a perfect octahedron with bond lengths obtained by averaging those from 5 structures of the compound found in the Cambridge Structure Database (Entries CEJYEV, GIJVEA, IDOZAX, QEZXUO and VUYBAS). The stereochemistry of ferrocyanide was strongly restrained, using the highest energy constants encountered for amino-acid residues. During refinement with CNS (Brunger et al., 1998), the structure of ferrocyanide remained very close from ideality, and the absence of noticeable peaks in the difference map justified *a posteriori* the choice of strong restraints. Only reflections common to both the LD and HD data sets were retained. Additionally, 89 of these reflections were rejected on the ground that the difference between their intensities in each data set was aberrant (Ursby et al., 1997). Once a model had been obtained for the mixed-valence state, this structure was refined against the HD data set with 200 cycles of positional and 100 cycles of individual B-factor refinement.

**Composite data sets.**

The technique of composite data sets (Berglund et al., 2002) was used to reconstruct a "low dose" (LD) and a "high dose" (HD) data sets (Figure 4b), which, based on spectroscopic results, can be confidently assigned to the oxidized and reduced states of the active center iron. The LD data set corresponded to a total exposure of 7 sunder a 17-times attenuated beam (first arrow on Figure 4b), whereas the HD data set corresponded to a total exposure of ~ 70 sunder the un-attenuated beam (second arrow on Figure 4b). LD and HD data sets were obtained from 12 pairs of subsets of 7 images, collected at different locations of the crystal, translating the latter by 100 μm and rotating it by 6º between subsequent subsets so as to scan the reciprocal space entirely. Statistics for the composite data sets are shown in Table 1. The resolution was limited to 1.7 Å due to the need for a reduced exposure for the LD data set, but the crystal could diffract to higher resolution. The absorbed dose for the LD data set was very small (8.7 $10^3$ Gy, *i.e.* 0.05 % of the Henderson limit, (Henderson, 1990) and it was still limited for the HD data set (1.5 $10^6$ Gy, *i.e.* 7.8 % of the Henderson limit), in which no detectable degradation of the diffractions patterns could be noticed. Based on an estimation of the crystal content, the approximate number of photons absorbed per unit cell was evaluated with the program Raddose (Murray et al., 2004) to be 0.001 and 0.2 for the LD and HD data sets, respectively. Assuming that a single absorbed X-ray photon may generate a cascade of up to ~ 500 e$^-$, ~ 0.5 e$^-$ and ~ 100 e$^-$ may eventually be guided to electron sinks such as metal centers in the LD and HD cases, respectively. Therefore, with ~ 16 iron atoms per unit cell (omitting the depleted center I), these findings are fully consistent with negligible and almost full photo-reduction of the active center iron in the LD and HD data sets, respectively.

**Steady-State Measurement of the Oxidation of SOR by $O_2^{\bullet-}$ in the presence of ferrocyanide**

The kinetics of the oxidation of SOR by $O_2^{\bullet-}$, generated by the xanthine-xanthine oxidase system, was followed by spectrophotometry (Lombard et al., 2000a) in the absence or in the presence of various molar equivalents of potassium hexacyanoferrate(II) (Sigma). Assays were performed at 25°C in 1 ml of reaction buffer (50 mM Tris/HCl, pH 7.6) containing 17 μM of SOR, 400 μM xanthine, 500 U catalase. The reaction was initiated by adding 0.09 U of xanthine oxidase. SOR oxidation was followed at 640 nm, and the kinetics was found to be linear for at least 5 min.

## <u>ACKNOWLEDGMENTS</u>

The online microspectrophotometer was bought and installed with support from the ESRF, the EMBL, and a Royal Society Equipment Grant to LMB, Oxford. We thank Raimond Ravelli, Martin Weik and Elspeth Garman for helping us with this device and for fruitful discussions on radiation damage issues. Marc Fontecave and Sine Larsen are acknowledged for discussions on iron coordination and careful reading of the manuscript. This work was supported by grants from the Commissariat à l'Energie Atomique and the Centre National de la Recherche Scientifique. V.A. was supported by an ESRF Traineeship. F.P.M.H acknowledges a post-doctoral fellowship from the CEA.

**FIGURES CAPTIONS**

**Figure 1. Three-dimensional structure of *D. baarsii* SOR.** The color changes from blue to cyan (respectively from yellow to red) from the N-terminal to the C-terminal ends of monomer A (respectively monomer B). Iron atoms are represented as dark-green spheres, with residues coordinating them shown as gray sticks. His49, His69, His75, His119 and Cys116 coordinate the Center II iron(II), whereas Cys10, Cys13, Cys29 and Cys30 coordinate the Centre I iron(III). Domains I and II are linked by a loop (residues 37-45) and a short $3_{10}$ helix (46-49). A calcium ion, depicted in pink, is found at the interface between the two monomers, with hexa-coordination to Cys88 and Thr90 from both monomers, and to two water molecules located near the symmetry axis. Prepared with PyMOL (DeLano Scientific, San Carlos, CA, USA).

**Figure 2. Details of the SOR active site.** Final $2F_o$-$F_c$ electron density map, contoured at 1.0 σ, superimposed onto a ball-and-stick representation of the active site. Center II iron and chloride are represented as dark-green and lime-green spheres, respectively. Iron coordination is figured by purple dotted lines. The chloride anion only binds tightly in monomer A, where Lys48 is highly ordered. In monomer B, a crystalline contact partially disorders Lys48, thus disrupting the water network at this location, and leading to alternate positions for the chloride ion, as suggested by an elongated electron density (not shown). Figures 2, 5b, 5c, 6 and 7 were prepared with Bobscript (Esnouf, 1999) and Raster 3D (Merritt et al., 1997).

**Figure 3. Offline absorbance spectra recorded on a SOR crystal co-crystallized with ferricyanide**. **(a)** Before X-ray data collection (black line). In grey, an absorbance spectrum on a native SOR crystal is shown as a reference, which does not show the absorption band at 650 nm. **(b)** After exposure to X-rays for 90 s (recorded on the same sample).

**Figure 4. Online difference absorbance spectra recorded on a SOR crystal co-crystallized with ferricyanide. (a)** Series of spectra collected during a 300 s X-ray exposure. The color changes from blue to red as the X-ray dose accumulate from 0 to ~ 6.4 $10^6$ Gy. The peak developing at ~ 540 nm is assigned to absorption by trapped solvated electrons. The peak developing below 450 nm is assigned to the reduction of disulfides bridges. **(b)** Decay of the 650 nm absorption band upon X-ray exposure. Experimental data are shown in black and fitting by a bi-exponential kinetic model is in red. The two arrows indicate the total exposure time used for the two composite data sets. (**inset**) Photograph of a SOR crystal bleached by the X-ray beam of ESRF station ID14-4 for 90 s.

**Figure 5. Plugging of the SOR active center by ferrocyanide. (a)** and **(b)** Electrostatic surface representations of the SOR active site, respectively in the absence, and in the presence of ferrocyanide, contoured from negative (red) to positive (blue) potentials, and prepared with GRASP (Nicholls et al., 1991). The negatively-charged ligand complements the protein's positively-charged active site. **(c)** Details of the interactions between ferrocyanide and SOR (stereo view). The cyanide moieties are interleaved between the equatorial histidines, stabilized by both hydrogen bonds (red dotted lines) and van der Waals interactions (thin black dotted lines). Coordination of the Center II iron is figured by purple dotted lines. **(d)** Side-view showing that ferrocyanide binds the active site in a bent geometry. **(e)** Same view of the chloride-bound native structure. Chloride binds to the active site in a strikingly similar way to ferrocyanide, with water molecules of the hydration sphere replacing cyanides for hydrogen bonding and van der Waals interactions with the protein.

**Figure 6. Electron density difference maps revealing structural modifications of SOR upon X-ray induced photo-reduction. (a)** Map contoured at +/- 4.0 σ (red: negative, green: positive) superimposed on SOR Center II (monomer B). **(b)** Map

contoured at +/- 4.5 σ superimposed on SOR Center I (monomer A). The iron atom of Center I (shown as an orange dot), is 60 % depleted, causing the formation of disulfide bonds between the cysteines normally involved in iron coordination. The positive and negative densities unambiguously illustrate the breakage of these disulfide bridges upon photo-reduction.

**Figure 7. Initial velocity of the oxidation of SOR Center II by superoxide in the presence of different molar equivalents of ferrocyanide.** (●) wild-type SOR and (❏) E47A mutant SOR. Representative traces of the kinetics of the oxidation of the Center II iron in wild-type SOR are shown in the inset, followed at 25 °C by the increase of absorbance at 640 nm. Data are normalized with 100% initial velocity obtained in the absence of ferrocyanide, being 0.0075 $OD_{640nm}$/min for both the wild-type and E47A mutant SORs.

**TABLES**

**Table 1**. **Data reduction, model refinement and structure validation.**

| | SOR$_{E47A}$ | SOR$_{E47A}$–Fe(CN)$_6$ mixed-valence | SOR$_{E47A}$–Fe(CN)$_6$ reduced |
|---|---|---|---|
| Resolution [Å] | 1.15 | 1.7 | 1.7 |
| R$_{sym}$[a] [%] | 5.1 (40.6)[b] | 8.3[c] (32.2) | 8.6[c] (38.0) |
| Completeness [%] | 99.0 (98.1) | 98.3 (94.1) | 97.8 (93.9) |
| Redundancy | 6.1 | 2.9 (2.9) | 2.9 (2.8) |
| <I/σ(I)> | 13.6 (2.8) | 12.3 (3.7) | 11.7 (3.0) |
| Unique reflections | 93,621 | 26,217 | 26,101 |
| Observations/parameters ratio | 4.2 | 2.8[d] | 2.8[d] |
| R$_{cryst}$[e] [%] | 13.6 | 20.6 | 20.7 |
| R$_{free}$[f] [%] | 17.8 | 24.9 | 25.0 |
| Estimated maximal error [Å][g] | 0.037 | 0.063 | 0.066 |
| Diffraction-data precision indicator [Å][g] | 0.039 | 0.149 | 0.151 |
| Overall G-factors[h] | -0.11 | 0.20 | 0.20 |
| Rmsd[i] bonds (Å) | 0.016 | 0.005 | 0.005 |
| Rmsd angles | 0.034 Å | 1.45º | 1.44º |
| Average B factor (Å$^2$) | 23.3 | 18.1 | 18.6 |

[a]$R_{sym} = \Sigma_j\Sigma_h \mid I_{h,j} - <I_h> \mid /\Sigma_j\Sigma_h I_{h,j}$

[b]Numbers in parentheses refer to the highest resolution shell (10 % of all reflections).

[c]$R_{merge}$ = 7.1% between the mixed-valence and reduced data sets (the cell volume expanses by 0.24 %).

[d]Refinement is performed with 25836 reflections common to the two data sets.

[e]$R_{cryst} = \Sigma_h \mid \mid F_{obs}(h) \mid - \mid F_{calc}(h) \mid \mid /\Sigma_h \mid F_{obs}(h) \mid$

[f]$R_{free}$ is calculated from a set of 5% randomly-selected reflections that were excluded from refinement.

[g]Errors in atomic positions calculated by SFCHECK from the various methods of Cruickshank (Vaguine et al., 1999). The expected maximal error is estimated from the properties of the electron-density map and DPI is derived from R$_{free}$, resolution, completeness, number of atoms and number of reflections.

[h]G-factor is the overall measure of structure quality from PROCHECK (Laskowski et al., 1993).

[i]rmsd: root mean square deviation.

**Table 2**. **SOR active site geometry.**

| PDB access code | XXX | | XXX | | XXX | |
|---|---|---|---|---|---|---|
| **Structure name** | *D. baarsii* SOR$_{E47A, Fe+2}$ | | *D. baarsii* SOR$_{E47A, Fe+3}$-Fe$_{+2}$(CN)$_6$ | | *D. baarsii* SOR$_{E47A, Fe+2}$-Fe$_{+2}$(CN)$_6$ | |
| **Redox state and spin state of the active centre iron** | (reduced, S = 2) | | (mixed-valence, S = 5/2) | | (reduced, S = 2) | |
| | Monomer A[a] | Monomer B | Monomer A | Monomer B[b] | Monomer A | Monomer B[b] |
| **Distances Fe-N$_{His}$ [Å]** | | | | | | |
| His49, N$_\epsilon$ | 2.193 (0.016)[c] | **2.167 (0.016)[c]** | **2.25** | **2.24** | **2.25 (0.00)[d]** | **2.24 (0.00)[d]** |
| His69, N$_\epsilon$ | 2.150 (0.014)[c] | **2.168 (0.019)[c]** | **2.18** | **2.18** | **2.25 (+0.07)[d, e]** | **2.18 (0.00)[d, e]** |
| His75, N$_\epsilon$ | 2.248 (0.015)[c] | **2.202 (0.016)[c]** | **2.24** | **2.23** | **2.24 (0.00)[d, e]** | **2.28 (+0.05)[d, e]** |
| His119, N$_\delta$ | 2.153 (0.013)[c] | **2.190 (0.019)[c]** | **2.13** | **2.13** | **2.18 (+0.05)[d]** | **2.17 (+0.04)[d]** |
| **Transhystidyl distances [Å]** | | | | | | |
| His49-His75 | 4.342 | **4.244** | **4.45** | **4.44** | **4.45 (0.00)** | **4.49 (+0.05)** |
| His69-His119 | 4.192 | **4.241** | **4.31** | **4.32** | **4.43 (+0.12)** | **4.35 (+0.03)** |
| Compression | 0.15 | **0.00** | **0.14** | **0.12** | **0.02** | **0.14** |
| **Distance Fe-S$_\gamma$ (Å)** | **2.364 (0.005)** | **2.359 (0.005)** | **2.64** | 2.54 | **2.70 (+0.06)** | 2.63 (+0.09) |
| **Distance Fe-Equat. Plane (Å)** | **0.47** | **0.51** | **0.16** | 0.11 | **0.14** | 0.11 |
| **Angle (bridging CN)-Fe (°)** | - | - | **141.4** | 149.0 | **142.3** | 150.3 |
| **Distance (bridging CN)-Fe (Å)** | - | - | **2.36** | 2.18 | **2.43 (+0.07)** | 2.22 (+0.04) |
| **Active site volume[f] (Å$^3$)** | - | - | **15.77** | 14.95 | **16.62** | 15.64 |

[a]in monomer A, due to crystal packing effects (in particular a H-bond between HisA119 and LysA55 from a symmetry-related molecule), distances in light-faced characters are considered less significant.

[b]in monomer B, binding of Fe(CN)$_6$ is altered by a crystalline contact between LysB48 and GluA100 from a symmetry-related molecule. Thus distances in light-faced characters are considered less significant.

[c]numbers in parenthesis indicate the estimated standard deviation computed by Shelx after full-matrix least-square refinement.

[d]numbers in parenthesis indicate the variations relative to the mixed-valence structure.

[e]motions of His69 and His75 might be affected by long-range crystalline contacts.

[f]active site volume is the volume of the octahedron formed by the six ligands of the active centre iron. The volume expansion measured in monomer A and B are 5.3 % and 4.6 %, respectively.

**Figure 1**

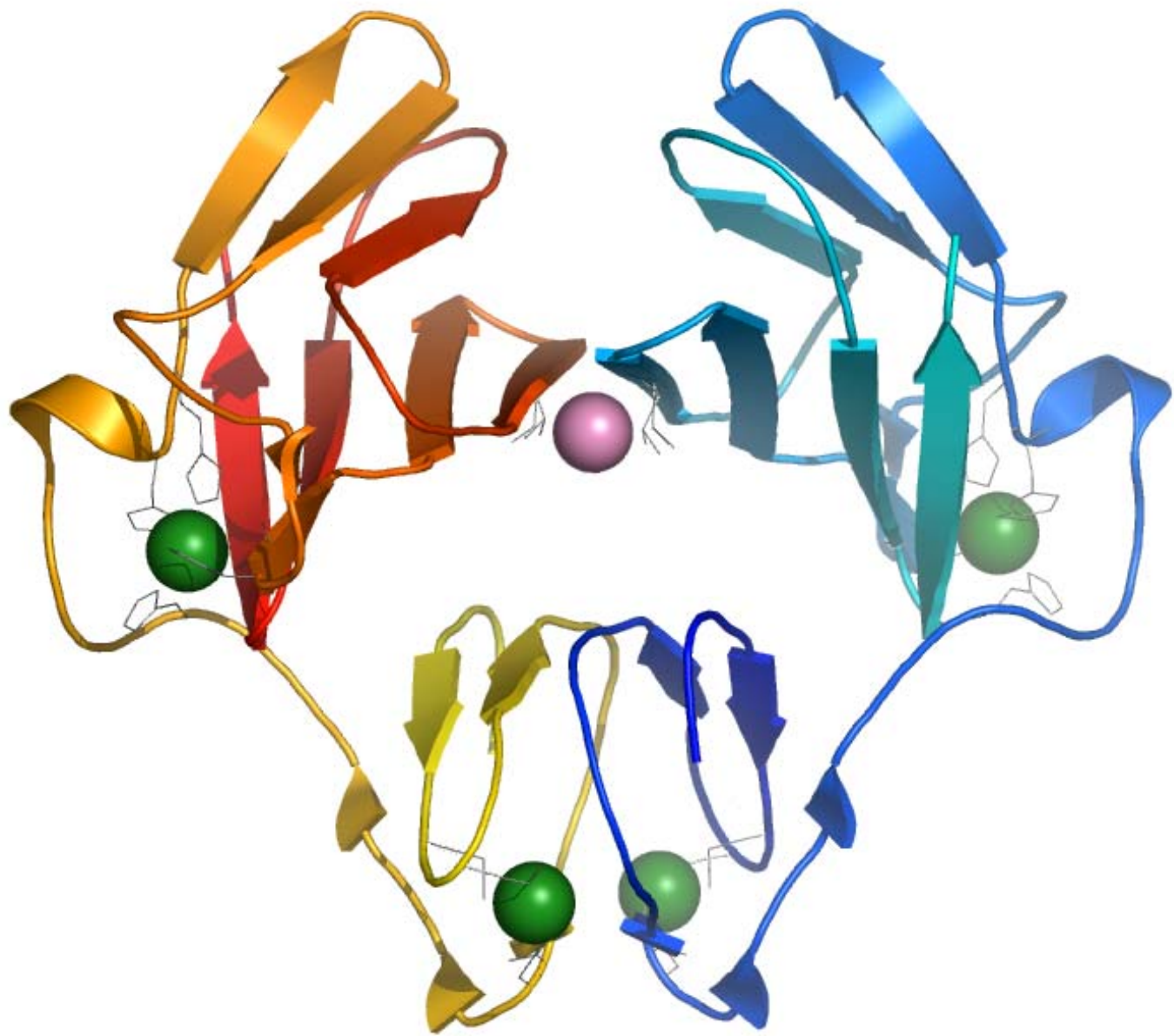

**Figure 2**

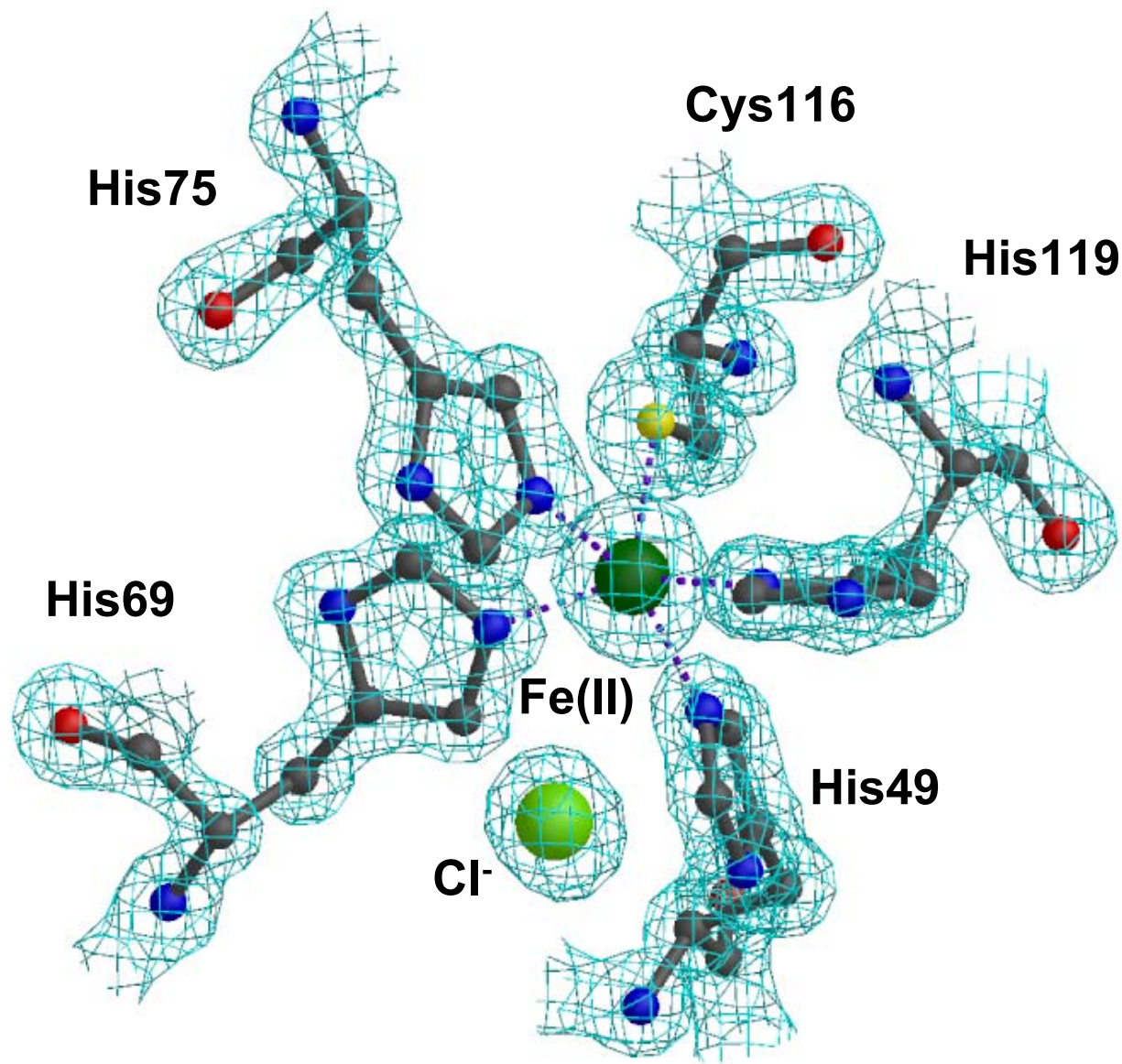

**Figure 3**

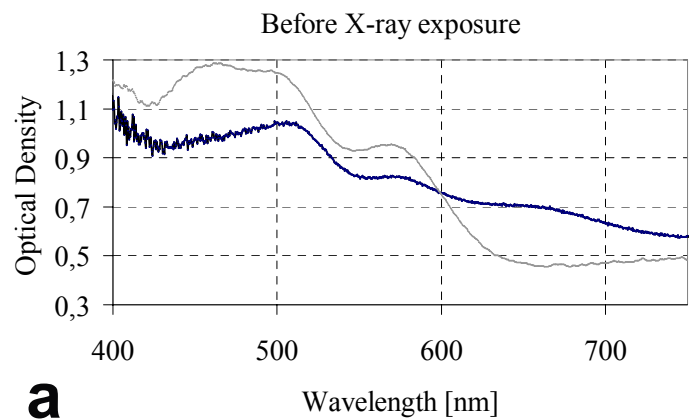

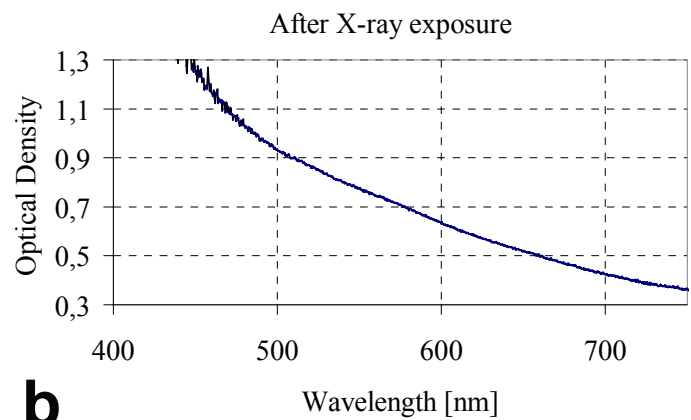

a

b

**Figure 4**

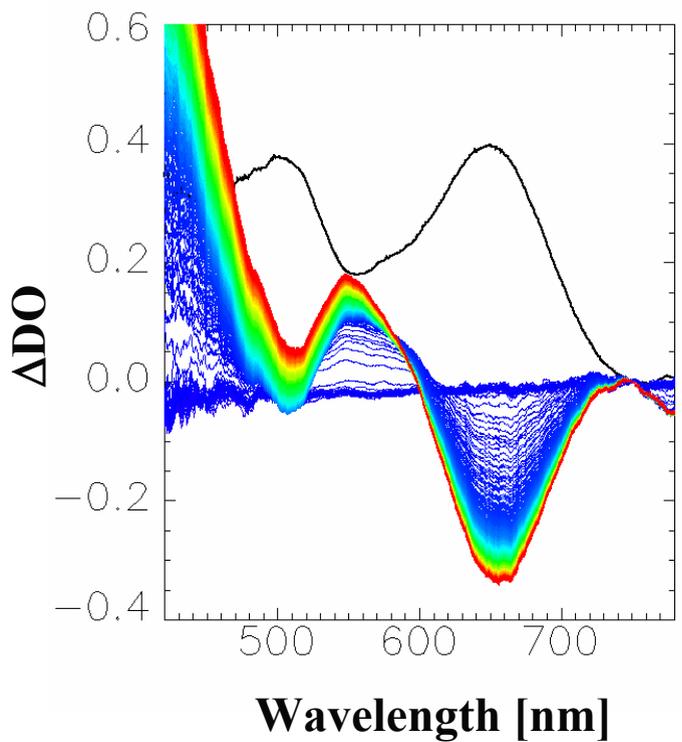

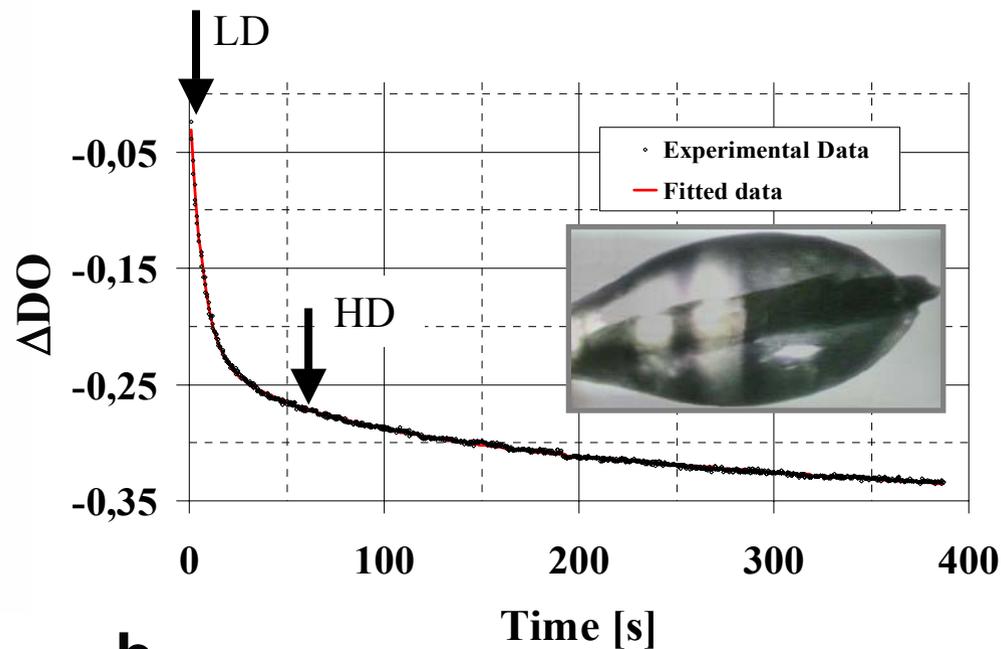

a

b

**Figure 5**

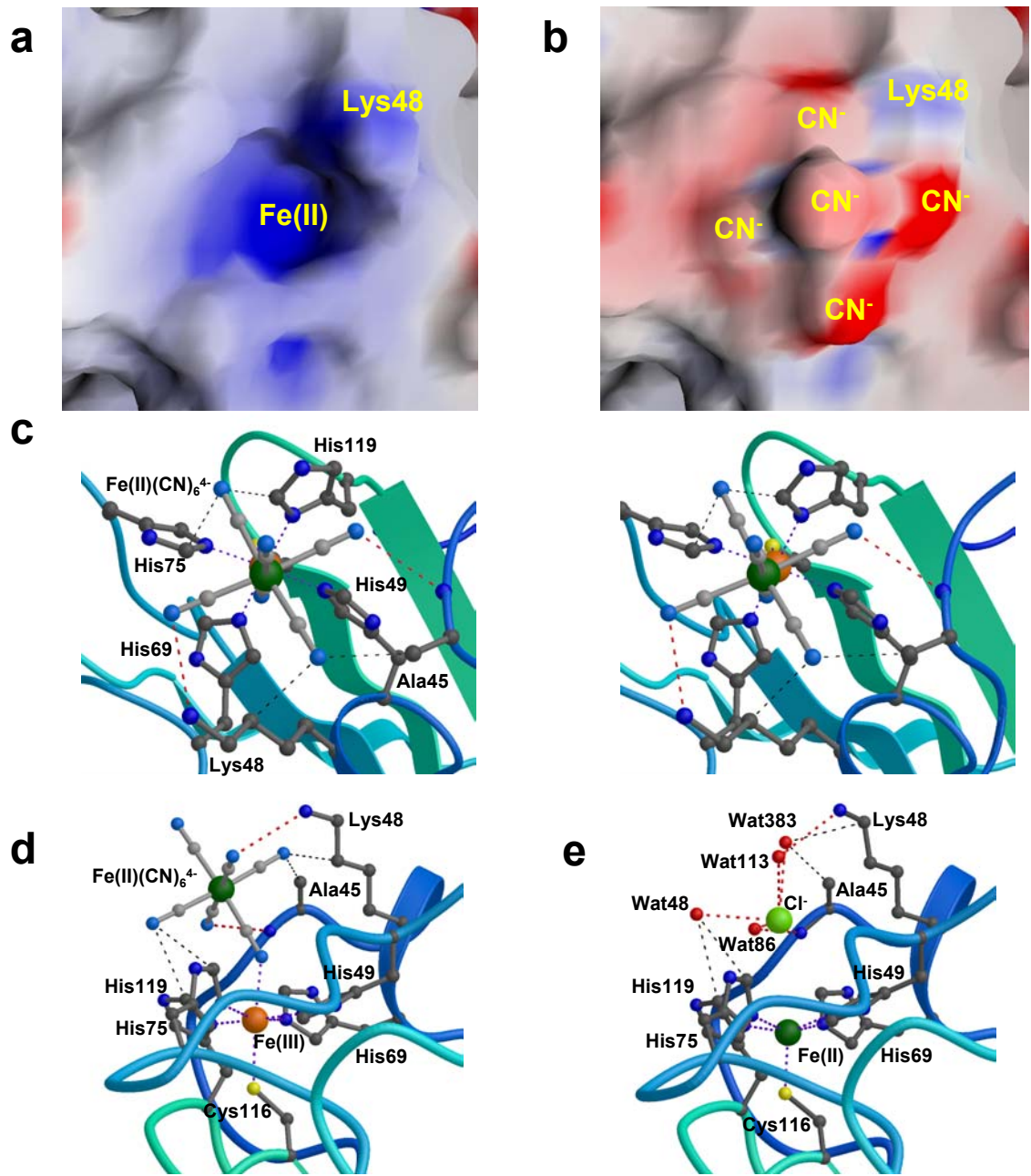

**Figure 6**

a

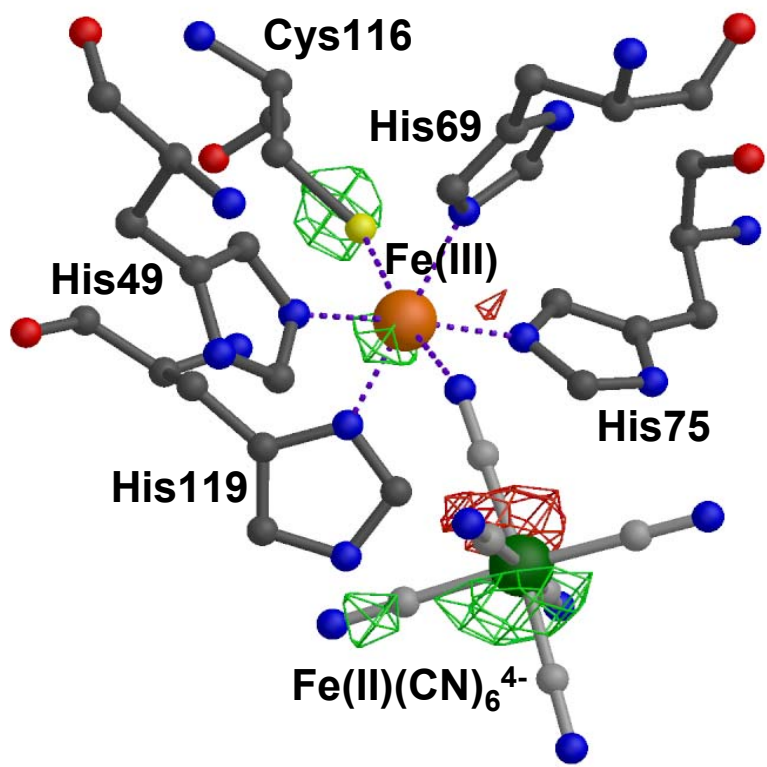

Cys116

His69

Fe(III)

His49

His75

His119

Fe(II)(CN)$_6^{4-}$

b

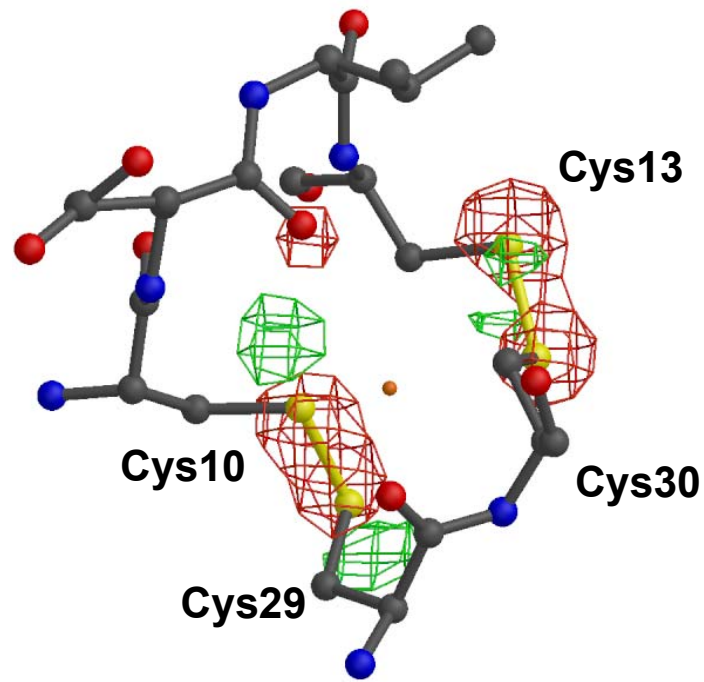

Cys13

Cys10

Cys30

Cys29

**Figure 7**

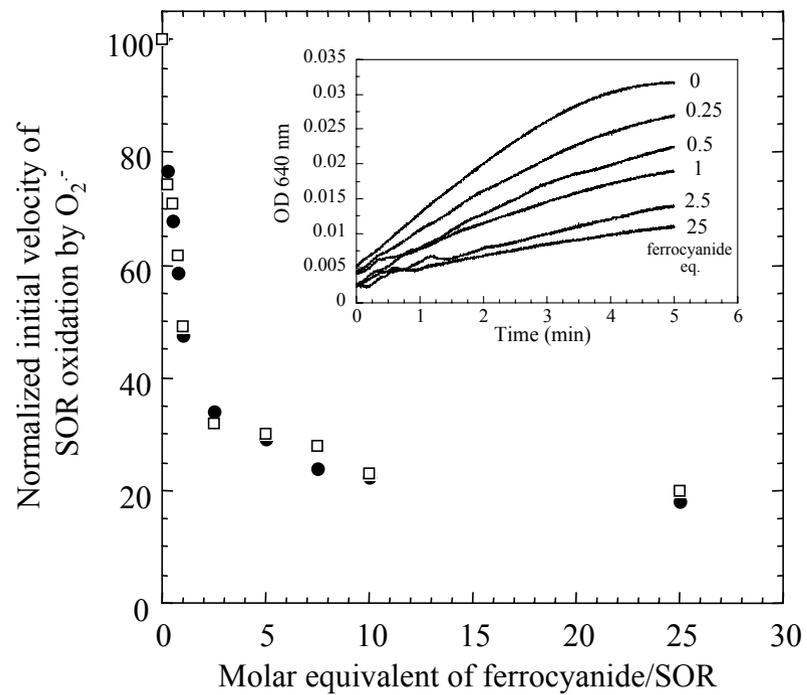